\documentclass[twocolumn,showpacs,preprintnumbers,amsmath,amssymb]{revtex4}

\usepackage{graphicx}
\usepackage{dcolumn}
\usepackage{bm}

\begin{document}


\title{Efficient community-based control strategies in adaptive networks}

\author{Hui Yang}
\affiliation{ Web Sciences Center, University of Electronic
Science and Technology of China, Chengdu 610054, P. R. China}

\author{Ming Tang}\email{tangminghuang521@hotmail.com}
\affiliation{ Web Sciences Center, University of Electronic
Science and Technology of China, Chengdu 610054, P. R. China}

\author{Hai-Feng Zhang}\email{haifeng3@mail.ustc.edu.cn}
\affiliation{School of Mathematical Science, Anhui University,
Hefei 230601, P. R. China}

\date{\today}

\begin{abstract}

Most researches on adaptive networks mainly concentrate on the properties of steady state,
but neglect transient dynamics. In this study, we pay attention to the
emergence of community structures in transient process and the effects
of community-based control strategies on epidemic spreading.
First, by normalizing modularity $Q$, we investigate the evolution of community
structures during the transient process, and find that very strong community
structures are induced by rewiring mechanism in the early stage of epidemic spreading,
which remarkably delays the outbreaks of epidemic.
Then we study the effects of control strategies started from different stages on the prevalence.
Both immunization and quarantine strategies indicate that it is not
``the earlier, the better" for the implementing of control measures.
And the optimal control effect is obtained if control measures
can be efficiently implemented in the period of strong community structure.
For immunization strategy, immunizing the
S nodes on SI links and immunizing S nodes randomly have
similar control effects. Yet for quarantine strategy, quarantining
the I nodes on SI links can yield far better effects than
quarantining I nodes randomly. More significantly, community-based quarantine strategy
plays more efficient performance than community-based immunization strategy.
This study may shed new lights on
the forecast and the prevention of epidemic among human population.

\end{abstract}


\maketitle

In various real-world systems, the structures are constantly changing with
the states of the network and vice versa~\cite{peter1}.
For instance, frequent traffic congestions on a road may lead to the building of new roads,
while the new roads will influence the traffic flow on some roads~\cite{review1}.
These phenomena are characterized by the existence of a feedback loop
between the dynamics in networks and the dynamics of networks,
and the networks with such a feedback loop are called coevolutionary or adaptive
network~\cite{review1,review2}.
Until recently, a variety of adaptive networks have been studied,
such as biological networks~\cite{schaper}, chemical
networks~\cite{jain}, ecological systems~\cite{dieckmann1,drossel} and
technological networks~\cite{Scire}. And many interesting
phenomena are found, including the robust self-organization phenomenon in
biological nervous systems~\cite{bornholdt1,bornholdt2}, the promotion
cooperation in evolutionary game~\cite{co-game3,co-game4}, the
emergence of community structure in opinion
spreading~\cite{co-opinion4}, etc.

From epidemiological viewpoint, when an infectious disease appears
in a population, human self-protection behaviors can significantly
change the predicted course of epidemics~\cite{funk,fenichel}.
In the extreme, susceptible people may break their contacts with
infected partners. This will significantly alter the structure
of the contact network, thus influencing the pathway of epidemic spreading.
Gross~\emph{et al.}~first studied the dynamics of
susceptible-infected-susceptible(SIS) model in adaptive
networks, and found that different epidemic transmission
rates and rewiring rates can lead to fascinating phenomena,
like bistability and circles~\cite{gross1}. Shaw and Schwartz~\cite{shaw1}
considered a susceptible-infected-recovery-susceptible (SIRS) model
in adaptive networks, and came into similar phenomena. Subsequently,
Marceau \emph{et al.} developed a more precise analytical method for
the adaptive networks in the framework of the model of Gross~\cite{marceau}.
In addition, quite a few studies showed that adaptive networks
can effectively change the dynamics of epidemic~\cite{gross2,segbroeck}
and contact switching is an effective control strategy for epidemic outbreaks~\cite{zanette1,zanette2,shaw2}.

The previous studies of epidemic spreading in adaptive
networks mainly focus on steady state while ignoring transient dynamics.
However, the understanding of transient dynamics would
help to put forward efficient and timely control strategy. In this paper, we
look into the adaptive SIS model of Gross
\emph{et al.}~\cite{gross1} again, and find that very strong community
structures are induced by rewiring mechanism in the early stage of epidemic spreading.
Then two community-based control strategies are proposed,
and a counter-intuitive conclusion is discovered:
it is not ``the earlier, the better" for the implementing of control measures.

The model considers SIS dynamics in a random network with fixed $N$ nodes
and $K$ undirected links~\cite{gross1}, where a node may be in S or I state.
For every SI link, I node infects S node with fixed probability $p$ per time step. And I node recovers from the disease with
probability $r$, becoming susceptible again. Meanwhile, with probability $w$ for every SI link, S node
rewires the link to a randomly selected S node. Multiple connections and self-connections
are excluded. Results presented in the following are
for $N=10^4$ and $K=10^5$.

\begin{figure}
\begin{center}
\includegraphics[height=60mm,width=90mm]{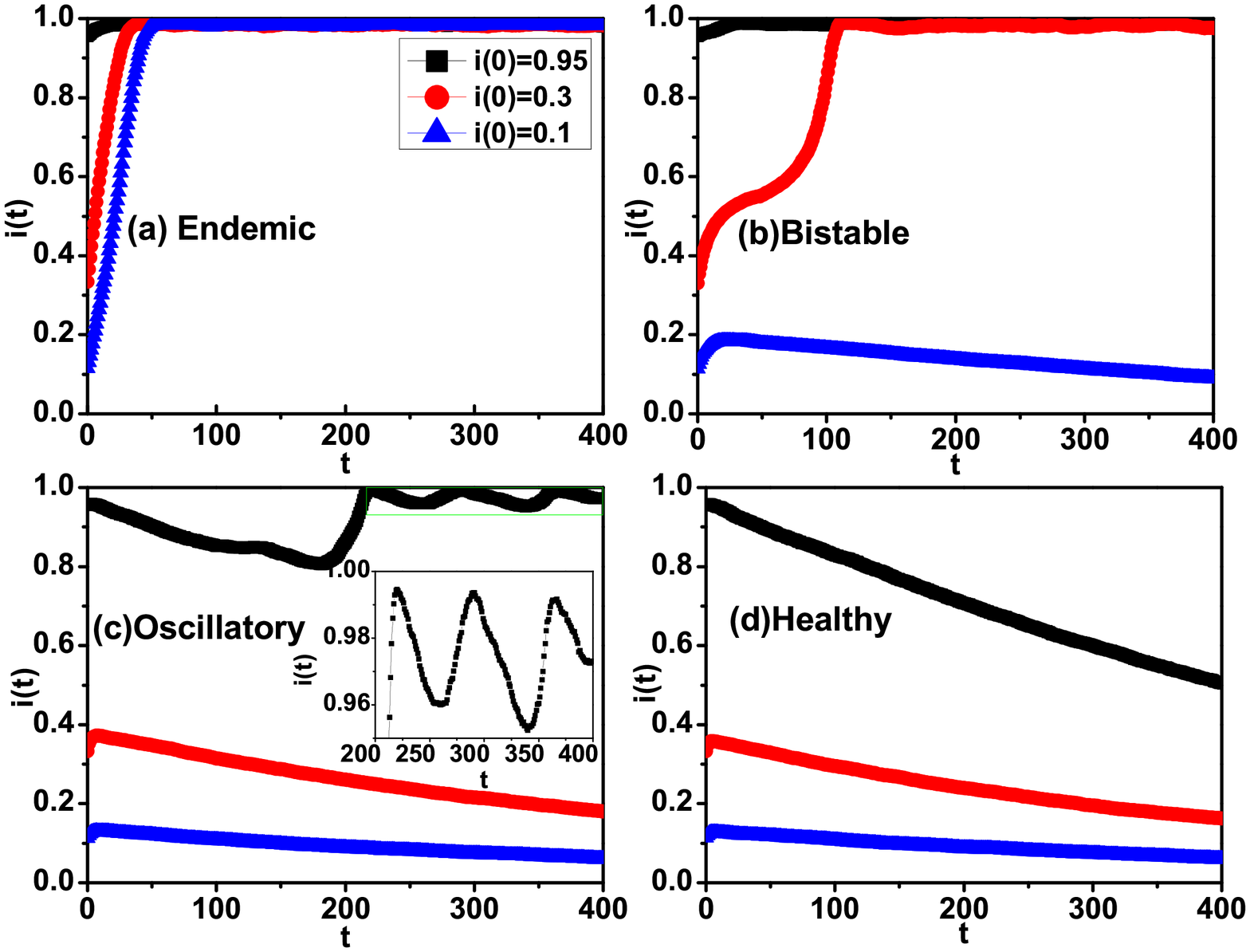}
\caption{(Color online) The density of infected $i(t)$ (i. e., the prevalence) versus time t for
different initial densities of infected $i(0)$, where ``squares", ``circles" and ``triangles"
denote the results of $i(0)=0.95,0.3,0.1$, respectively.
Four figures represent different phases: (a) Endemic
state, $w=0.1$; (b) Bistable state, $w=0.3$; (c) Oscillatory
state, $w=0.6$; (d) Healthy state, $w=0.7$.
Here $p=0.008,r=0.002$. Inset: the magnification of the black line labelled
green region.}\label{fig1}
\end{center}
\end{figure}

For different values of $w$ and $p$, the system
can be divided into four different phases~\cite{gross1}: endemic state,
bistable state, oscillatory state, and healthy state (see Fig.~\ref{fig1}).
Note, in bistable state and oscillatory state, different initial densities
of infected $i(0)$ can yield completely different results. In Fig.~\ref{fig1}(b),
for $i(0)=0.1$, large rewiring rate $w$
can effectively isolate I nodes and prevent S nodes from
infection, leading to the dying out of epidemic. For $i(0)=0.3$,
the rewiring behavior can lower the outbreak velocity at the early stage, but
can not prevent the outbreak of epidemic. For $i(0)=0.95$,
the impact of rewiring behavior is negligible due to the huge
number of I nodes. When $w$ increases to 0.6, there is an oscillatory state for large $i(0)=0.95$ in
Fig.~1(c). Though large rewiring rate $w$ can prevent the prevalence to some extent,
the gradually increasing S nodes will form a
giant and tight cluster with the smaller threshold, which causes re-outbreak in S cluster.
This process will repeat so many times, and an oscillatory phenomenon appears.

From above analysis, we know that the rewiring behavior can
lead to two separated but internally compact clusters.
Naturally, we can define the community structure of the
adaptive network according to the different states of nodes~\cite{state-community}.
Strong community structure implies that both $S$ and $I$ nodes are more
likely to connect to nodes with the same state and the number of
SI links is relatively small, while weak community structure indicates that S and
I nodes are mixed more fully in the network. Network modularity $Q$, a popular
evaluating indicator in measuring community structure~\cite{newman}, is defined as
\begin{equation}\label{1}
Q=\sum_{s=1}^{C}[\frac{l_s}{L}-(\frac{d_s}{2L})^2],
\end{equation}
where $l_s$ and $d_s$ represent the number of intra-links and the sum of degrees of the nodes in community $s$
respectively, $L$ denotes the number of links in the
network, and $C$ is the number of communities.  Here $0\leq
Q\leq1$, the larger $Q$ is, the stronger community structure is.

\begin{figure}
\begin{center}
\includegraphics[height=60mm,width=90mm]{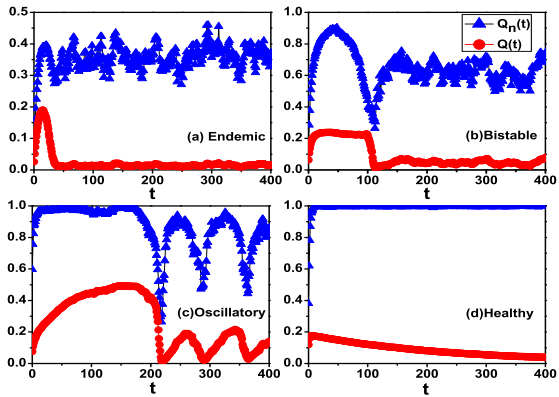}
\caption{(Color online) The time evolution of
 $Q(t)$ (``circles") and $Q_n(t)$ (``triangles") under different conditions. (a)
$i(0)=0.3, w=0.1$; (b) $i(0)=0.3, w=0.3$; (c) $i(0)=0.95, w=0.6$;
(d) $i(0)=0.3, w=0.7$. When $i(0)$ is small, oscillatory state is
the same as the healthy state; Oscillation occurs only when $i(0)$
is large. So we choose $i(0)=0.95$ in oscillatory state. Here
$p=0.008, r=0.002$.}\label{fig2}
\end{center}
\end{figure}

Fig.~\ref{fig2} shows the time evolution of $Q(t)$ in different phases.
One can find that, sometimes, the indicator $Q(t)$ cannot
well characterize the community strength of the network. For example,
when the system is in healthy state, most of SI links can be
rapidly broken by the large rewiring rate $w$, which may bring about the
complete separation of S and I clusters. Hence the community
structure in healthy state should be more obvious than in the other states.
Yet from Fig.~\ref{fig2}(d), one can find that $Q(t)$ is very low,
especially for large time steps. It should be noted that such
difference comes from the shortcomings of modularity $Q$. Specifically,
for $C=2$, Eq.~(\ref{1}) is expanded as
\begin{equation}\label{2}
Q=\sum_{s=1}^{2}[\frac{l_s}{L}-(\frac{d_s}{2L})^2]=1-\frac{l_{12}}{L}-(\frac{d_1}{2L})^2-(\frac{d_2}{2L})^2,
\end{equation}
where $l_{12}$ represents the amount of inter-links between
community $1$ and $2$. When the network is connected randomly, $l_{12}=(d_1d_2)/2L$, thus
$Q=0$; When $l_{12}=0$ and $l_1 =l_2$, $Q$ reaches the maximum, i. e., $Q=0.5$.
Therefore $Q$ can only range from $0$ to $0.5$. Moreover, the large
difference between intra-link number of community $1$ and $2$,
e. g., $d_1\gg d_2$, also induces small Q even when $l_{12}$ is small.
For instance, community A(B) is a complete graph with 50(10) nodes
and there is only one link between the two communities.
The modularity of such a community network is
$Q=1-1/1271-(2451/2542)^2-(91/2542)^2\approx0.068$.

Above all, $Q$ is not an accurate index to characterize the community structure of the network with two
communities. To address this shortcoming, like Ref.~\cite{kashtan}, the
normalized $Q_n$ is defined as
\begin{equation}\label{3}
Q_n=\frac{Q-Q_{rand}}{Q_{max}-Q_{rand}},
\end{equation}
where $Q_{rand}$ corresponds to random network with the same degree sequence,
and $Q_{max}$ is the modularity of the network without inter-community links, i. e., $l_{12}=0$.
After this standardization, $Q_n$ can range from $0$ to $1$.

The normalized $Q_n$, in comparison to $Q(t)$,
can reflect the community structure in the adaptive network more accurately. In endemic state,
as shown in Fig.~\ref{fig2}~(a), though small rewiring rate $w$ can't
prevent the outbreak of epidemic, it can also induce a certain degree of community structure.
In bistable state, Fig.~\ref{fig2}~(b) demonstrates that $w=0.3$ can give rise to
the form of large and tight S cluster in the early stage ($t\leq50$).
And subsequent re-outbreak in S cluster ($50<t\leq110$) results in the weakening of community strength.
When $t>110$, the community strength of the network will keep a stable value.
In oscillatory state, $w=0.6$ can isolate I nodes quickly, hence the
community strength becomes large rapidly. But with S cluster becomes larger
and tighter, the epidemic threshold becomes smaller. Finally, epidemic prevails
in S cluster at some point, such as $t\approx200$ in Fig.~2(c), resulting in the weakening of community structure.
Then this process will repeat again and again.
In healthy state, the very large $w$ will separate S and I cluster completely and
rapidly, thus the community structure is increasingly strong in
the early stage and keeps $Q_n\approx1$ after S and I clusters are separated (see
Fig.~3(d)).
\begin{figure}
\begin{center}
\includegraphics[height=60mm,width=90mm]{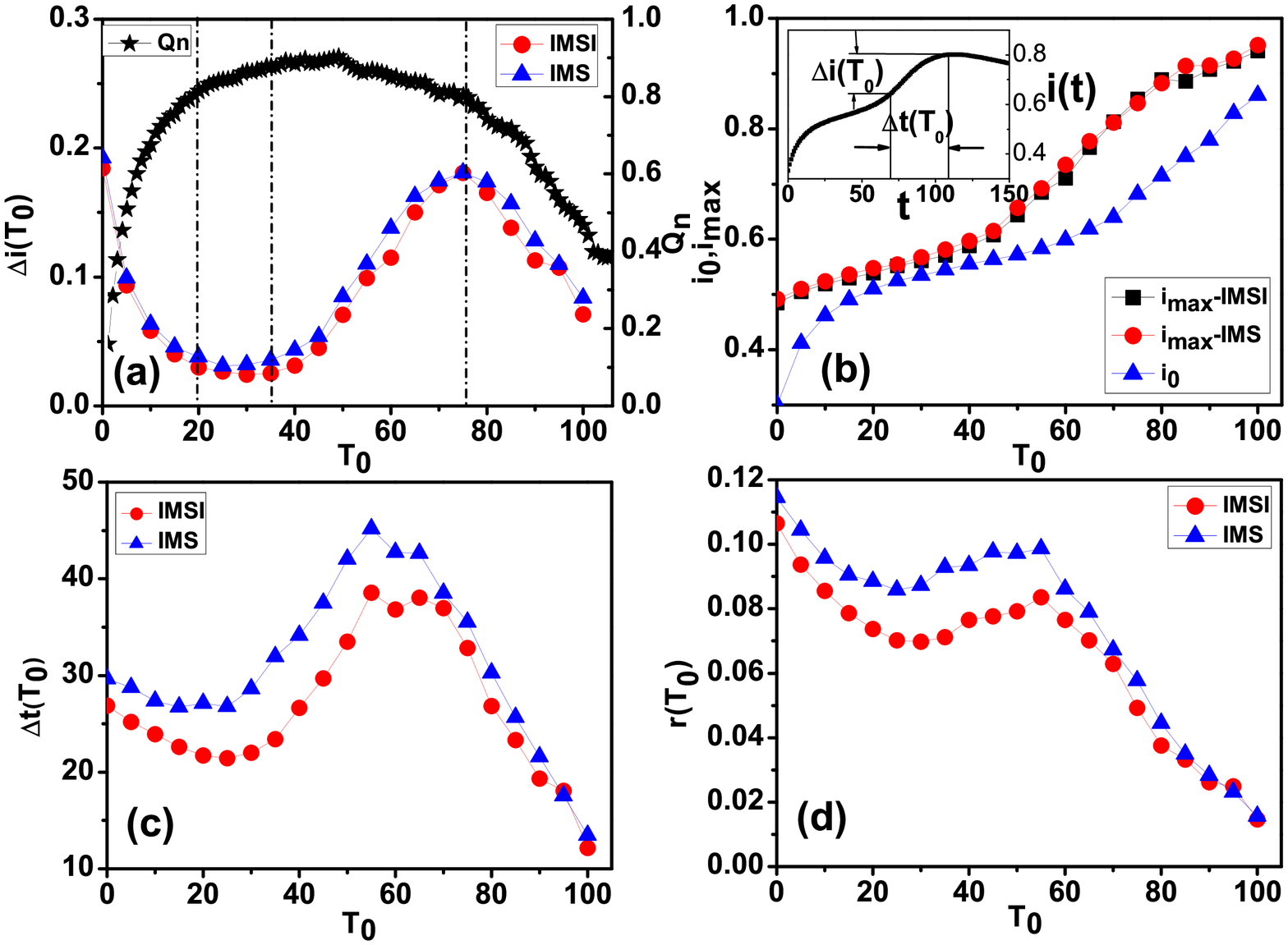}
\caption{(Color online) The effects of different immunization strategies vs the
starting time of immunization $T_0$. (a) $\Delta i(T_0)$, (b) $i_0, i_{max}$, (c) $\Delta t(T_0)$, (d) $r(T_0)$.
Here $i(0)=0.3, w=0.3, r=0.002, p=0.008$, and $f=0.008$. ``Stars" in (a) represent the time evolution
of $Q_n(t)$. Inset of (b) shows the definition of $\Delta i(T_0)$ and $\Delta t(T_0)$.
Each point represents an average over 200 realizations.
}\label{fig3}
\end{center}
\end{figure}

From the above, S nodes and I nodes can be
divided into two loosely coupled but internally tight communities
due to the rewiring mechanism in the transient process.
One question is whether this property is helpful for the control of disease.
In this viewpoint, we study the impacts of adaptive community structure on the control of disease,
and pursue efficient control strategies. Look back to transient processes
in the four phases of the system. In endemic state, disease prevails
so rapidly that it is difficult to detect the community structure and control
the epidemic; In healthy state, there is no need for controlling disease;
In oscillatory state, the oscillatory phenomenon occurs only when $i(0)$ is very large,
such as $i(0)>0.9$, but it is somewhat unrealistic. In bistable state,
small $i(0)$ can arouse the outbreak, and there is a very strong community structure for a long time.
Therefore, we only focus on the bistable state.

In the transient process, the strength of community structure is always changing.
To undestand the effects of adaptive community structures on the
prevention of epidemic, we compare the control effects started from different stages.
As we all know, immunization and quarantine are two basic measures to control epidemic spreading.
Here we consider immunization strategy and quarantine strategy, respectively.

\textbf{Immunization strategy}---immunizing a fraction $f$ of S nodes per
time step. Two approaches to choose S nodes are compared:
(1) randomly choose S nodes from the network (labelled IMR);
(2) randomly select S nodes from SI links (labelled IMSI).
To evaluate the effect of the control strategy, we let $T_0$ be the
starting time of immunization (quarantine), $\Delta i(T_0)$ be the difference
between the density of infected $i(T_0)$ at time $T_0$ and the maximal density of infected $i_{max}(T_0)$
that can reach after immunization (quarantine), i. e., $\Delta
i(T_0)=i_{max}(T_0)-i(T_0)$, $\Delta t(T_0)$ denote the time interval of this process,
and $r(T_0)$ denote the total percentage of immunized (quarantined) nodes in $\Delta t(T_0)$
(see inset of Fig.~\ref{fig3} (b)).

In Fig.~\ref{fig3}, the results of IMR and IMSI strategies are compared.
Figs.~\ref{fig3}~(a), (c) and (d) show that
these curves follow similar trends. What is more interesting is that three parameters
are minimal when $T_0\in [20,35]$. So the optimal starting time of immunization
is $T_0\in [20,35]$ which is somewhat against our intuition --
the earlier immunization starts, the better control effect is.
To explicitly explain such a phenomenon, the time window of $T_0$ is
divided into four regions according to the trend of $\Delta i(T_0)$
(see Fig.~\ref{fig3}~(a)): $\Delta i(T_0)$ decreases for $T_0\in [0,20]$,
keeps stable for $T_0\in [20,35]$, increases gradually for
$T_0\in [35,75]$ and drops again for $T_0\in [75,100]$.

For $T_0\in [0,20]$, the strength of community structure
and the density of infected $i(t)$ increases rapidly.
When the community strength is weak, immunizing few S nodes
can not effectively break the bridges between S cluster and I cluster.
Thus, the increase of prevalence is large, e. g., $\Delta i(T_0=0)\approx0.2$.
As the delaying of the starting time $T_0$, the community structure will be enhanced,
and thus IMSI strategy can break the connections between the two communities more thoroughly,
which makes the prevalence soon be brought under control.
Consequently, $\Delta i(T_0)$ decreases as $T_0$,
which is exactly opposite to traditional concept ``the earlier, the better".
Meanwhile, the epidemic can be inhibited faster than before
(i.e., smaller $\Delta t(T_0)$ in Fig.~3(c)) with immunizing
fewer S nodes on SI links (i.e., $r(T_0)$ in Fig.~3(d)).

For $T_0\in [20,35]$, the community structure is very strong
and the density of infected $i(t)$ increases slowly.
Owing to the very strong community structure,
immunizing S nodes on SI links can timely hold back the re-outbreak in S cluster,
which results in the minimum of $\Delta i(T_0)$.
In Figs.~\ref{fig3}~(c) and (d), $\Delta t(T_0)$ and $r(T_0)$ also
reach their minimum and keep stable.

For $T_0\in [35,70]$, the adaptive network also has very strong community structure.
However, it will certainly take some time to hold back the increase of prevalence,
e. g., $\Delta t(T_0)\approx20$ for $T_0=20$. Suppose immunization starts after $T_0=35$,
there is no enough time left for controlling, because tight S cluster is invaded gradually by
epidemic again which leads to the weakening of community structure at $t\approx50$ (see Fig.~\ref{fig1}(b)).
So re-outbreak is inevitable due to the lack of enough controlling time.
Therefore, the latter the starting immunization time $T_0$,
the worse the control effect of IMSI strategy.

For $T_0\in [70,100]$, the prevalence increases rapidly,
while the strength of community structure decreases dramatically.
Though the trends of these curves are the same as the region $T_0\in [0,20]$,
the reason is completely different. $i_{max}(T_0)\rightarrow 1$ for $T_0\in [70,100]$ (see
Fig.~3(b)), because control strategies no longer have any significance in this stage.
$\Delta i(T_0)\approx 1-i(T_0)$ thus decreases with $T_0$.

\begin{figure}
\begin{center}
\includegraphics[height=60mm,width=90mm]{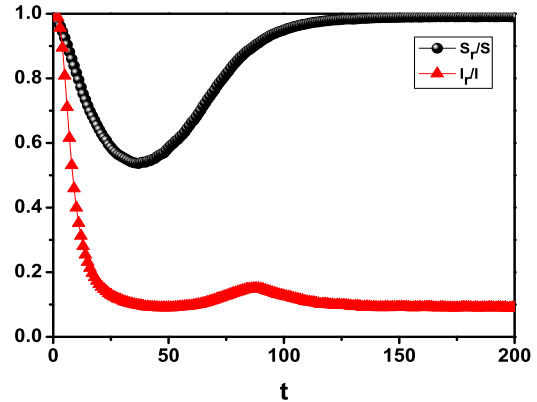}
\caption{(Color online) In bistable state, $S_r/S$ and $I_r/I$ vs time t, where
$S_r/S$ (``circles") represents the percentage of S nodes located on SI links
in all S nodes and $I_r/I$ (``triangles") represents the percentage of I nodes
located on SI links in all I nodes. Results are derived from an average over 200 realizations.}\label{fig4}
\end{center}
\end{figure}

Interestingly, in Fig.~\ref{fig3}, the impact of IMS strategy is
almost equal to IMSI strategy. To explain this reason,
the percentage of S nodes on SI links in all S nodes ($S_r/S$) is shown in Fig.~\ref{fig4}.
Although there is the strongest community structure at $t\approx50$,
the ratio $S_r/S\approx0.5$ is still relatively large.
Consequently, both IMS strategy and IMSI strategy have similar results.

\textbf{Quarantine strategy} -- quarantining a fraction $f$ of I nodes per time
step. Similar to immunization, two approaches to choose I nodes are also
considered: (1) randomly choose I nodes from the
network (labelled ISR); (2) randomly selected I nodes from SI links (labelled ISSI).

\begin{figure}
\begin{center}
\includegraphics[height=60mm,width=90mm]{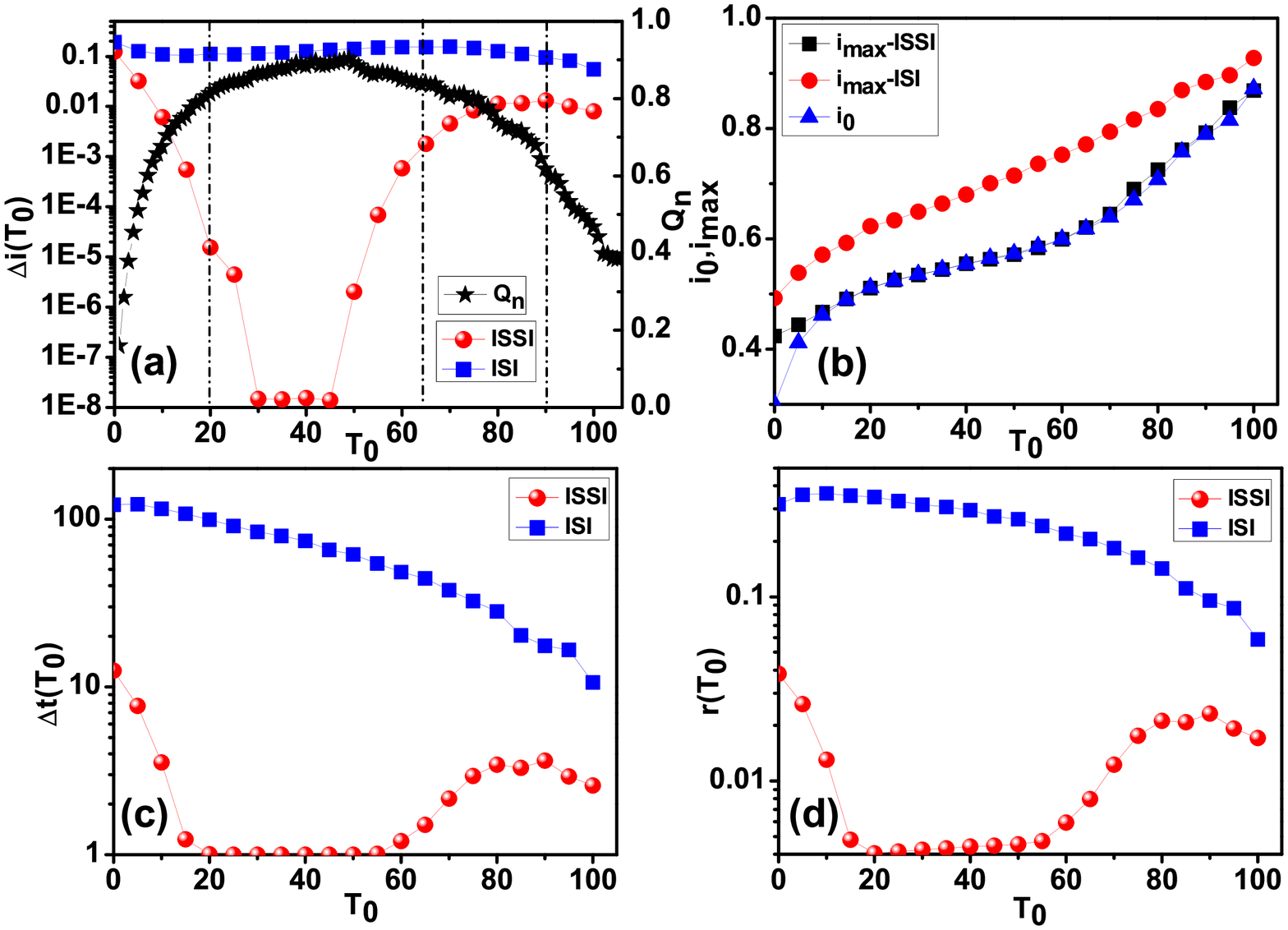}
\caption{(Color online) The effect of different quarantine strategies vs the starting
time of quarantine $T_0$. (a) $\Delta i(T_0)$, (b) $i_0, i_{max}$, (c) $\Delta t(T_0)$, (d) $r(T_0)$.
Here $i(0)=0.3, w=0.3, r=0.002, p=0.008$, and $f=0.008$. ``Stars" in (a) represent the time evolution
of $Q_n(t)$. Each point represents an average over 200 realizations.}\label{fig5}
\end{center}
\end{figure}

The effects of ISR strategy and ISSI strategy are compared in Fig.~\ref{fig5}. Unlike the
immunization strategies, the effect of ISR strategy is much worse
than ISSI strategy. As illustrated in Fig.~4, the ratio
$I_r/I$ is small, that is, most of infected nodes are not on the
SI links. As a result, ISR strategy can hardly pitch on
the I nodes on SI links, and the advantage of community structure
could not be well developed by ISR strategy. However, the effect of ISSI strategy is striking
since ISSI strategy can efficiently cut the pathways of disease to
the S cluster. Especially, $\Delta i(T_0)$ in
Fig.~\ref{fig5} (a), $\Delta t(T_0)$ in Fig.~\ref{fig5} (c) and $r(T_0)$ in Fig.~\ref{fig5} (d) reach the minimum
when $T_0\in [20,65]$, which corresponds to the time interval of strongest
community structure (Here, we should note that, compared with IMSI
strategy, ISSI strategy is more efficient in controlling
the epidemic and has larger optimal region since ISSI strategy can directly
cut more spreading pathways to the S cluster than
IMSI strategy). The reasons of the trends of these curves in Fig.~\ref{fig5}
are similar to the case of immunization strategy: In initial stage
($T_0\in [0,20]$), community strength increases with $T_0$.
Therefore, larger $T_0$ brings about better control effect.
Then it reaches the optimal region, i. e., $T_0\in [20,65]$,
in which the community structure is strongest. ISSI strategy can contain epidemic spreading rapidly.
So $\Delta i(T_0)$, $\Delta t(T_0)$ and $r(T_0)$ reach the minimum and keep stable.
When $T_0\in [65,90]$, the compact S cluster is invaded by disease again
and the community structure becomes weak gradually, so the effect of ISSI
strategy become poor with $T_0$. At last, $i_{max}(T_0)\approx1$ for $T_0>90$,
thus $\Delta i(T_0)$ decreases with $T_0$ again.

To sum up, we studied the properties
of community structures in the transient process of adaptive networks
by the standardized modularity. We found that different degrees
of community strength emerge from distinct rewiring conditions.
Especially in bistable state, the very strong community structure
can hold for a long period. In view of this, community-based
immunization strategies and quarantine strategies are studied thoroughly.
Because most S nodes are on SI links when epidemic prevails in an adaptive
network, random immunizing S nodes can give rise to
similar effects as immunizing S nodes on SI links.
Nevertheless, for quarantine strategy, quarantining I
nodes on SI links is significantly better than random quarantining
I nodes in the network, since the former can efficiently
cut the pathways of epidemic invading to the S clusters.
Both the study of immunization and quarantine strategies discover a
counter-intuitive conclusion: it is not ``the earlier, the better"
for the implementing of control measures. And the optimal control effect is
obtained if control measures can be efficiently implemented
in the period of strong community structure.
More significantly, community-based quarantine strategy
plays more efficient performance than community-based immunization strategy.

The prevalence of an infectious disease can
trigger behavioral responses of people trying to minimize the risk
of being infected. If so, further study on the control strategy in
adaptive networks has instructive significance. And we have done a
forward step along this line. This is helpful for controlling
adaptive dynamics in real world, such as epidemic and rumor spreading.

\acknowledgments
This work was jointly funded by the National Natural Science
Foundation of China (Grant Nos. 11105025, 11005001), China Postdoctoral
Science Foundation (Grant No. 20110491705), the Specialized Research
Fund for the Doctoral Program of Higher Education (Grant No. 20110185120021), and
the Fundamental Research Funds for the Central Universities (Grant No. ZYGX2011J056).


\end{document}